\documentstyle[12pt]{article}
\begin{document}
{\hskip 11.6cm} SNUTP 96-085\par
%\vspace{1ex}
%{\hskip 12.0cm} hep-ph/961mnn\\
%\vspace{1ex}
%{\hskip 12.0cm} August, 1996\\
\vspace{5ex}
\begin{center}        
{\LARGE \bf Heavy Baryon Masses in Large ${\rm N_c}$ HQET}\\
\vspace{7ex}
{\sc Chun Liu}\footnote{\it email: liuc@ctp.snu.ac.kr}\\
\vspace{3ex}     
{\it Center For Theoretical Physics, Seoul National University}\\
{\it Seoul, 151-742, Korea}\\

\vspace{10.0ex}
{\large \bf Abstract}\\
\vspace{4ex}
\begin{minipage}{130mm}
                                                                               
   We argue that in the large ${\rm N_c}$ HQET, the masses of the s-wave
low-spin heavy baryons equal to the heavy quark mass plus proton mass 
approximately.  To the subleading order, the heavy baryon mass 
${\rm 1/N_c}$ expansion not only has the same form, but also has the same
coefficients as that of the light baryon.  Based on this, numerical 
analysis is made.  
\par
\vspace{2.0cm}
\end{minipage}
\end{center}

\newpage
                                                                               
  Heavy baryons provide us testing ground for the Standard Model.  Those
containing a single heavy quark, like $\Lambda_c$, $\Lambda_b$, 
$\Sigma_c^{(*)}$ and $\Sigma_b^{(*)}$, can be studied within the heavy
quark effective theory (HQET) [1].  For complete calculations for them, 
some additional nonperturbative methods have to be used.  In this Letter,
we discuss the simple incorporation of large ${\rm N_c}$ [2] method in 
HQET. 
\par
\vspace{1.0cm}    
   HQET is an effective field theory of QCD in the heavy quark limit [1].
In a systematic manner, it fits the description for the heavy hadrons.
Under the heavy quark limit, there is no heavy quark pair production.  The 
large mass of the heavy quark which interacts with the light quark system
with typical energy ${\rm \Lambda_{QCD}}$, plays no role except for the 
total energy of the hadron.  With the velocity super-selection rule, the
heavy quark mass $m_Q$, which is defined perturbatively as the pole mass, 
can be removed by the field redefinition.  The heavy quark field $h_v$ is
defined by
\begin{equation}
P_+Q(x)=\exp (-im_Qv\cdot x)h_v(x)~,
\end{equation}
where $P_+=\frac{1}{2}(1+\not \!\!v)$.  To the leading order of $1/m_Q$,
the effective Lagrangian for the heavy quark is
\begin{equation}
{\cal L}_{\rm eff}=\bar{h}_viv\cdot Dh_v~.
\end{equation}
Besides the heavy quark symmetry [1], we note explicitly from Eq. (2) that
the heavy quark becomes effectively massless (modula $m_Q$).  The heavy 
hadron 
mass $M$ is expanded as 
\begin{equation}
M=m_Q+\bar{\Lambda}~,
\end{equation}
where $\bar{\Lambda}$ is the heavy hadron mass in the HQET, which is 
independent of the heavy quark flavors.  The quantity $\bar{\Lambda}$ cannot
be determined from the HQET further.  It is at this stage, we apply the
large ${\rm N_c}$ method.
\par
\vspace{1.0cm}
  As one of the most important and interesting method of nonperturbative
QCD, large ${\rm N_c}$ limit [2] is often applied in spite of the realistic 
${\rm N_c}=3$.  Nonperturbative properties of mesons can be observed from 
the analysis of the planar diagram, and baryons from the Hartree-Fock 
picture.  Recently, there are renewed interests in the large ${\rm N_c}$
application to baryons due to the work of Ref. [3] which shows that there
is a contracted SU(2$f$) light quark spin-flavor symmetry in the baryon 
sector, by 
combining the large ${\rm N_c}$ counting rules and the chiral Lagrangian.
Actually this symmetry can be directly derived in the Hartree-Fock
picture [4], or by other method [5].  Similar result was also obtained 
before [6].  Further applications of this spin-flavor symmetry to heavy 
baryons
are made by Jenkins [3] in discussing the baryon-pion couplings and the 
baryon hyperfine splittings. 
Interesting relations among the baryonic Isgur-Wise
functions are obtained in Refs. [7] as well as [8].  Masses of the heavy 
baryons with any finite number of heavy quarks are studied by ${\rm 1/N_c}$
expansion of QCD in Ref. [9]. 
\par
\vspace{1.0cm}
  Inspired by these approaches, we consider the HQET at the large ${\rm N_c}$
limit.  Physically, the heavy quark limit and the large ${\rm N_c}$ limit 
are non-commutative.  Different order of the limits corresponds to different
picture.  In the large ${\rm N_c}$ HQET, there is nothing new in the meson 
case.  So we discuss the heavy baryons.   
\par
\vspace{1.0cm}
  We argue that the mass of the s-wave low-spin heavy baryons in the HQET
$\bar{\Lambda}$ equals to the proton mass in the large ${\rm N_c}$ limit.
Let us continue thinking of the Hartree-Fock picture not in the full QCD,
but in the HQET.  The heavy baryons contain $({\rm N_c} -1)$ light quarks,
and one "massless" heavy quark.  The mass or the energy of the baryon is 
determined by the summation of the energies of individual quarks.  The 
kinetic energy of the heavy quark is typically $\Lambda_{\rm QCD}$ like
that of the light quark.  The interaction energy between the heavy quark 
and any of the light quarks is typically $\Lambda_{\rm QCD}/{\rm N_c}$.  
So the
interaction energy between the heavy quark and the whole light quark
system scales as $\Lambda_{\rm QCD}$.  However, the total interaction
energy of the light quark system itself scales as
${\rm N_c\Lambda_{QCD}}$.  In the limit ${\rm N_c}\rightarrow \infty$, 
the light quarks drown the heavy quark.  The energy of the heavy baryon is
determined by its light quark system.  This light quark system also 
dominates the proton in the large ${\rm N_c}$ limit.  Therefore we come
to the conclusion:  in the large ${\rm N_c}$ limit, the
masses of the s-wave low-spin heavy baryons defined in HQET equal to the 
proton mass. 
\par
\vspace{1.0cm}
  From the same logic as in last paragraph, we can easily deduce the 
results for the baryon-pion coupling constants.  These constants are 
also determined by the light quark system.  So they are the same for 
the light baryons and the heavy baryons.  And the heavy baryon also has the
light quark spin-flavor symmetry.  These results are obtained by 
Jenkins in Ref. [3].
\par
\vspace{1.0cm}
  Of course, all the results are subject to $1/{\rm N_c}$ corrections which
deserve more detailed considerations.
The correction violates the light quark spin-flavor symmetry.  Let us first 
discuss the spin symmetry violation in ${\bar \Lambda}$. 
The baryon mass can be written as
\begin{equation}
\bar{\Lambda} = N_c\Lambda_{\rm QCD} + c_1J^2_l/N_c~, 
\end{equation}
where $J_l$ is the angular momentum of the light quark system.  The mass
parameter $c_1$ is yet undetermined which is of order 
${\Lambda_{\rm QCD}}$.  The factor ${\rm N_c}$ should appear so as to keep
the ${\rm N_c}$ scaling for $\bar{\Lambda}$.  In the extreme case while
all the quark spins align in the same direction, $J^2_l\sim {\rm N_c}^2$.
Only by dividing a factor ${\rm N_c}$, has the term $\sim J^2_l$ in 
Eq. (4) the right ${\rm N_c}$ scaling.  Note this term is ${\rm 1/N_c^2}$
suppressed compared to ${\rm N_c}\Lambda_{\rm QCD}$.  On the other hand, 
the light baryon mass $m$ has the same form of ${\rm 1/N_c}$ expansion,
\begin{equation}
m = {\rm N_c}\Lambda_{\rm QCD} + \tilde{c_1} J^2/{\rm N_c}~, 
\end{equation}
where $J$ is the baryon spin.  Further, we argue in the following that 
\begin{equation}
c_1 = \tilde{c_1}~.
\end{equation}
Consider still the above extreme case, where in the mass ${\rm 1/N_c}$
expansion, the subleading term becomes a leading one, 
$J^2=\frac{\rm N_c}{2}(\frac{\rm N_c}{2}+1)$ and 
$J^2_l=\frac{{\rm N_c}^2-1}{4}$.
Because of the light quark dominance, we have $m=\bar{\Lambda}$ in the
limit ${\rm N_c}\rightarrow \infty$.  This immediately results in the
conclusion given by Eq. (6).  
\par
\vspace{1.0cm}
  Another lowest order ${\rm 1/N_c}$ effect lies in the light quark 
flavor symmetry breaking.  At the moment, we forget the spin symmetry
violation.  After including the baryons with strangeness number $-1$, the
masses for the heavy and light baryons can be expanded as
\begin{equation}
\begin{array}{lll}
\bar{\Lambda}& = & {\rm N_c}\Lambda_{\rm QCD} + c_2 (-S)~,\\
m & = & {\rm N_c}\Lambda_{\rm QCD}+ \tilde{c_2} (-S)~,
\end{array}
\end{equation}
respectively.  Where $S$ is the baryon strangeness number which can be 
0 or $-1$.  Again we will argue
\begin{equation}
c_2 = \tilde{c_2}~.
\end{equation}
In the expression (7), the spin symmetry is not violated.  The strange 
quark spin decouples from the strong interaction.  The only 
contribution of the strange quark mass to baryon masses is the strange
quark mass itself.  Therefore $c_2$ and $\tilde{c_2}$ are nothing but
the strange quark mass defined in the large ${\rm N_c}$ limit.  
To the order $1/{\rm N_c}$, terms like $I^2$ and $I\cdot J_{(l)}$ should
be included in the expansion (7).  However, in the realistic case,
$I=J$.  These terms can be effectively absorbed into the term $J^2$ in 
Eq. (4).
\par
\vspace{1.0cm}
  For a complete analysis of the heavy baryon masses, $1/m_Q$ 
corrections have to be considered.  To the order of $1/m_Q$, heavy 
baryon mass $M$ is expanded as
\begin{equation}
M=m_Q+\bar{\Lambda}-\frac{\lambda_1}{2m_Q}+\frac{2\lambda_2}{m_Q}
(S_Q\cdot J_l)~,
\end{equation}
where $S_Q$ is the heavy quark spin and 
\begin{equation}
\begin{array}{lll}
\lambda_1&=&<H(v)|\bar{h}_v(iD)^2h_v|H(v)>~,\\
2\lambda_2(S_Q\cdot J_l)&=&\displaystyle -\frac{1}{4}Z_Q<H(v)|
\bar{h}_vg\sigma
\cdot Gh_v|H(v)>~,
\end{array}
\end{equation}
with $Z_Q$ being the renormalization factor.  In the leading order 
${\rm 1/N_c}$, $\lambda_1$ scales as unity and is independent of
the light quark structure; $\lambda_2$ is vanishing.  These can be 
seen directly from the definition (10) with light quark spin-flavor
symmetry, and from the fact that $\lambda_2$ is zero for $\Lambda_Q$
baryon.  Therefore we arrive the following ${\rm 1/N_c}$ expansion
for $\lambda_1$ and $\lambda_2$, 
\begin{equation}
\begin{array}{lll}
\lambda_1&=&c'_0+c'_1J_l^2/{\rm N_c}^2+c'_2S/N_c~,\\
\lambda_2&=&c''(S_Q\cdot J_l)/{\rm N_c}+c''_2S/{\rm N_c}~.
\end{array}
\end{equation}
\par
\vspace{1.0cm}
  We perform the numerical analysis for the non-strange baryons in
the following.  The heavy baryon mass is presented in Eq. (9).  For
$\bar{\Lambda}$ and $m$, the ${\rm 1/N_c}$ expansions are given in
Eqs. (4) and (5) with $c_1=\tilde{c_1}$.  And for $\lambda_1$ and
$\lambda_2$ in Eq. (11) with $S=0$.  To be consistent, the
accuracy of the analysis is maitained  to the order of 
$\frac{\Lambda_{\rm QCD}^2}{m_Q{\rm N_c}}$ and 
$\frac{\Lambda_{\rm QCD}}{{\rm N_c^2}}$.  That means the term 
$c'_1$ in Eq. 
(11) is also neglected.  Formally the uncertainty will be due to
$1/m_Q^2$ and $1/N_c^3$ corrections which are about 10 MeV.  With
the measured masses of proton, neutron and $\Delta$, we obtain
${\rm N_c}\Lambda_{\rm QCD}=866$ MeV and $c_1=293$ MeV.  This gives
$\bar{\Lambda}_{\Lambda_Q}=866$ MeV and 
$\bar{\Lambda}_{\Sigma^{(*)}_Q}=1060$ MeV.  Although there is no data
for $c'_0$, the following quantity can be predicted with the 
theoretical accuracy of 10 MeV, 
\begin{equation}
\begin{array}{lll}
\frac{1}{3}(M_{\Sigma_c}+2M_{\Sigma^*_c})&=&M_{\Lambda_c}+
\bar{\Lambda}_{\Sigma^{(*)}_c}-\bar{\Lambda}_{\Lambda_c}\\
&=&2479~{\rm MeV}~.
\end{array}
\end{equation}
Similarly the corresponding quantity for bottom quark is predicted as
\begin{equation}
\frac{1}{3}(M_{\Sigma_b}+2M_{\Sigma^*_b})=5835\pm50~{\rm MeV}~.
\end{equation}
Eq. (12) shows that the recent proposed $\Sigma^{(*)}_c$ masses in a 
new interpretion [10] of heavy baryon spectrum are in 100 MeV deviation
from our result.  It also implies that $M_{\Sigma^*_c}=2492$ MeV by 
taking 
$M_{\Sigma_c}=2453$ MeV.  Our numerical analysis actually is the same
as that in Ref. [9].
\par
\vspace{1.0cm}
  Comparing with Ref. [9], what are the different points of this paper?
We began with the HQET which gives a clear physical picture for heavy 
baryons, and emphasized the heavy baryon mass in HQET $\bar{\Lambda}$
is at the order of proton mass.  Then we showed that the next to leading
order ${\rm 1/N_c}$ expansions of $\bar{\Lambda}$ and the light baryon
mass not only have the same form, but also have the same coefficients.
These points cannot be taken for granted in large ${\rm N_c}$ HQET.
They justifies some of the numerical analysis of Ref. [9].
\par
\vspace{2.0cm}

   The author would like to thank M. Kim, S. Kim and P. Ko 
for helpful discussions.

\end{document}